\documentclass{osa-article}

\journal{oe}

\articletype{Research Article}

\usepackage{lineno}

\begin{document}

\title{Valley Hall edge solitons in a photonic graphene}

\author{Qian Tang,\authormark{1} Boquan Ren,\authormark{2} Victor O. Kompanets,\authormark{3} Yaroslav V. Kartashov,\authormark{3} Yongdong Li,\authormark{2} and Yiqi Zhang\authormark{2,*}}

\address{\authormark{1}Ministry of Education Key Laboratory for Nonequilibrium Synthesis and Modulation of Condensed Matter, Shaanxi Province Key Laboratory of Quantum Information and Quantum Optoelectronic Devices, School of Physics, Xi'an Jiaotong University, Xi'an 710049, China\\
\authormark{2}Key Laboratory for Physical Electronics and Devices of the Ministry of Education \& Shaanxi Key Lab of Information Photonic Technique, School of Electronic Science and Engineering, Xi'an Jiaotong University, Xi'an 710049, China\\
\authormark{3}Institute of Spectroscopy, Russian Academy of Sciences, Troitsk, Moscow, 108840, Russia}

\email{\authormark{*}zhangyiqi@xjtu.edu.cn} 



\begin{abstract}
We predict the existence and study properties of the valley Hall edge solitons in a composite photonic graphene with a domain wall between two honeycomb lattices with broken inversion symmetry. Inversion symmetry in our system is broken due to detuning introduced into constituent sublattices of the honeycomb structure. We show that nonlinear valley Hall edge states with sufficiently high amplitude bifurcating from the linear valley Hall edge state supported by the domain wall, can split into sets of bright spots due to development of the modulational instability, and that such an instability is a precursor for the formation of topological bright valley Hall edge solitons localized due to nonlinear self-action and travelling along the domain wall over large distances. Topological protection of the valley Hall edge solitons is demonstrated by modeling their passage through sharp corners of the $\Omega$-shaped domain wall.
\end{abstract}

\section{Introduction}

Topological insulators are considered as a new state of matter due to their unique physical properties.
In solid-state physics, where such insulators were introduced initially,
and where they are still under very active investigation, electronic topological insulators posses forbidden gap in the bulk,
just like conventional insulators, but allow conductance on their surface, even in the presence of considerable defects and disorder \cite{hasan.rmp.82.3045.2010,qi.rmp.83.1057.2011}.
The origin of this effect is connected with the existence of the edge states in the topological gap opening in the dispersion diagram in the momentum space.
These topologically protected entities are localized at the edge between two topologically distinct materials.
The notion of topological insulators over the last decade was extended to diverse branches of physics,
in particular to various wave systems.
Among them are photonic systems \cite{rechtsman.nature.496.196.2013, haldane.prl.100.013904.2008, wang.nature.461.772.2009,lindner.np.7.490.2011,hafezi.np.7.907.2011,stuetzer.nature.560.461.2018, yang.nature.565.622.2019,mukherjee.science.368.856.2020,maczewsky.science.370.701.2020,yang.light.9.128.2020},
acoustic \cite{yang.prl.114.114301.2015, peng.nc.7.13368.2016,he.np.12.1124.2016,lu.np.13.369.2017,zhang.cp.1.97.2018,ma.nrp.1.281.2019},
mechanical systems \cite{susstrunk.science.349.47.2015, huber.np.12.621.2016},
ultra-cold atoms \cite{goldman.pnas.110.6736.2013,jotzu.nature.515.237.2014},
polaritons in microcavities \cite{nalitov.prl.114.116401.2015,jean.np.11.651.2017,klembt.nature.562.552.2018},
and electrical circuits \cite{albert.prl.114.173902.2015,hadad.ne.1.178.2018,imhof.np.14.925.2018,olekhno.nc.11.1436.2020,helbig.np.16.747.2020,li.nsr.8.nwaa1192.2021}.
The progress in ``topological photonics'', rapidly growing into independent research area,
has been summarized in several recent reviews \cite{lu.np.8.821.2014,ozawa.rmp.91.015006.2019,kim.lsa.9.130.2020,smirnova.apr.7.021306.2020,ota.nano.9.547.2020,leykam.nano.9.4473.2020,segev.nano.10.425.2021,parto.nano.10.403.2021,wang.fo.13.50.2020,liu.col.19.052602.2021} covering various aspects of light propagation in these unusual materials.
Upon the development of topological photonics, it has been realized that nonlinear effects may be very important for the formation, manipulation, and control of the topological edge states.
Many interesting phenomena mediated by the nonlinearity have been discovered in topological insulators.
They include various bistability effects for edge states in pumped dissipative systems \cite{kartashov.prl.119.253904.2017, zhang.lpr.13.1900198.2019,zhang.pra.99.053836.2019},
modulational instabilities of the nonlinear edge states \cite{lumer.pra.94.021801.2016, kartashov.optica.3.1228.2016}, stabilization of operation of topological lasers due to nonlinear gain saturation \cite{harari.science.359.eaar4003.2018, bandres.science.359.eaar4005.2018,dikopoltsev.science.373.1514.2021,bahari.science.358.636.2017,kartashov.prl.122.083902.2019,zeng.nature.578.246.2020,zhong.lpr.14.2000001.2020,gong.acs.7.2089.2020},
nonlinearity-induced topological transitions \cite{maczewsky.science.370.701.2020},
as well as rich variety of solitonic effects \cite{Kartashov.nrp.1.185.20219,malomed.rjp.64.106.2019,mihalache.rrp.73.403.2021}, including the formation of self-sustained localized states in the bulk of topological insulators \cite{lumer.prl.111.243905.2013,mukherjee.science.368.856.2020},
nonlinear vortices \cite{bleu.nc.9.3991.2018}, topological edge solitons \cite{leykam.prl.117.143901.2016,ablowitz.pra.96.043868.2017,gulevich.sr.7.1780.2017,li.prb.97.081103.2018,smirnova.lpr.13.1900223.2019, zhang.prl.123.254103.2019,ivanov.acs.7.735.2020,ivanov.ol.45.1459.2020,ivanov.ol.45.2271.2020,ivanov.pra.103.053507.2021,zhong.ap.3.056001.2021,smirnova.prr.3.043027.2021},
or nonlinearity-induced higher-order topological phases \cite{zangeneh.prl.123.053902.2019}, to name a few.

Topological edge solitons were introduced as hybrid states that are affected by both the topological nature of the system and the nonlinear self-action.
Their investigation was mostly limited to polaritonic systems with external magnetic field \cite{kartashov.optica.3.1228.2016, gulevich.sr.7.1780.2017,li.prb.97.081103.2018,zhang.pra.99.053836.2019} and to waveguiding systems with longitudinal refractive index modulations \cite{leykam.prl.117.143901.2016,ivanov.acs.7.735.2020,ivanov.pra.103.053507.2021} serving to break time-reversal symmetry of the system.
At the same time, it is known that the appearance of topological edge states in valley Hall systems does not require time-reversal symmetry breaking and is associated instead with breakup of the inversion symmetry of the system [the word ``valley'' is associated here with specific features (presence of the local extrema) of bands of corresponding systems: for example, when inversion symmetry of the underlying honeycomb lattice is broken by detuning of two constituent sublattices, the gap opens between former Dirac points and local extrema in two upper bands develop that are called  valleys]. The latter setting therefore can be realized without using external magnetic fields or longitudinal system modulations, that are always associated with losses.
Even though valley Hall edge solitons were considered previously in sophisticated lattice geometries possessing type-II Dirac cones in the spectrum \cite{zhong.ap.3.056001.2021},
the specific structure of the underlying lattice did not allow illustration of their topological protection.
Such states, bifurcating from linear topological edge states at Bloch momenta yielding appropriate sign of the group velocity dispersion, and their topological protection so far were not considered at the domain walls between usual detuned honeycomb lattices (used in the majority of experiments on linear valley Hall edge states), which are much easier for experimental implementation.

In this paper, we report on valley Hall edge solitons forming at the domain wall in conventional honeycomb waveguide array (a photonic graphene).
We study properties of the linear and nonlinear edge states at such domain walls and present long-living topological edge solitons that demonstrate topological protection upon passage through sharp bends of the domain wall.
Our results suggest experimentally straightforward approach to implementation of such states.

\begin{figure}[htpb]
\centering
\includegraphics[width=0.5\columnwidth]{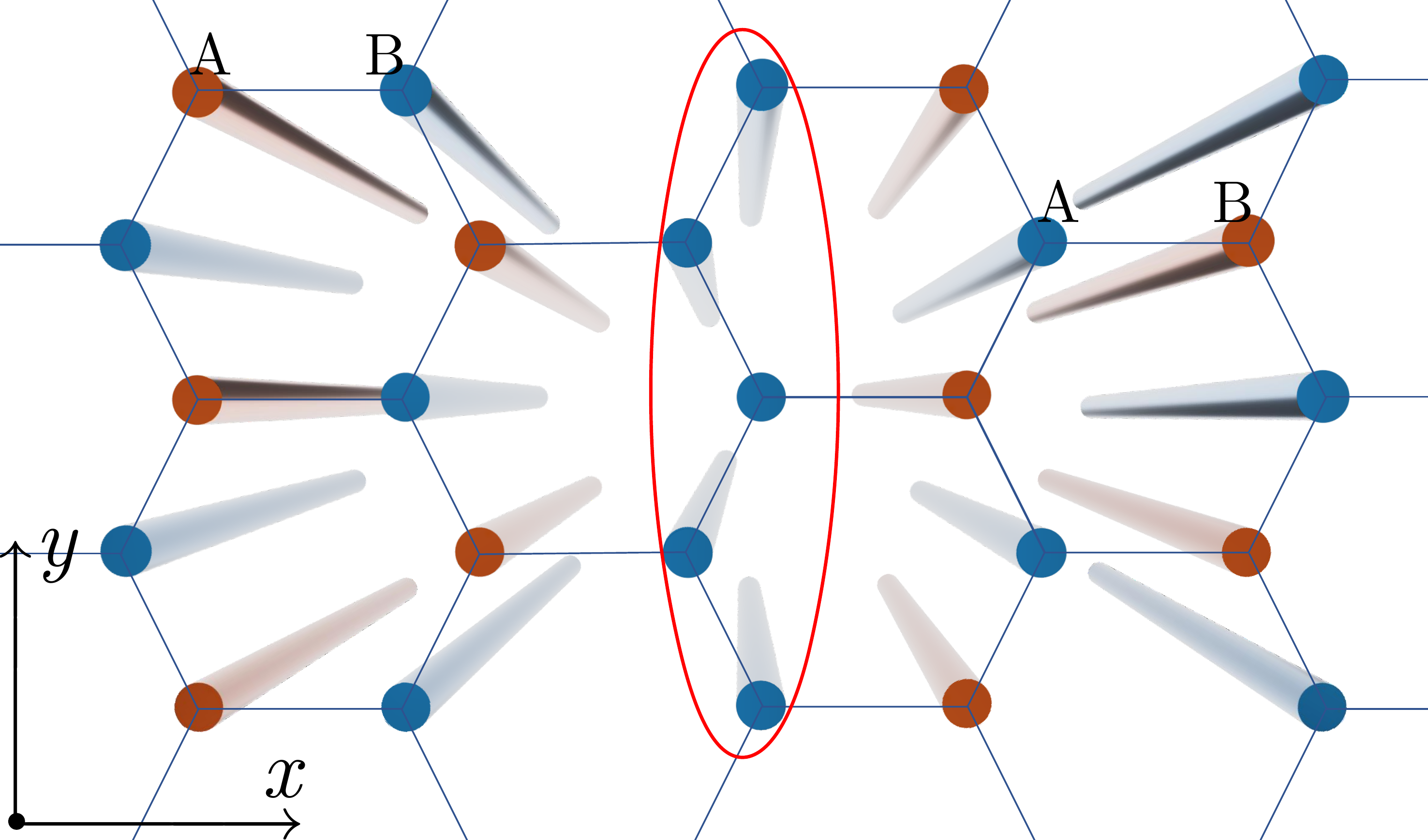}
\caption{Schematic configuration of honeycomb waveguide array with a domain wall. Two sublattices are distinguished by different colors. Domain wall between two honeycomb arrays with different detunings is highlighted by the red ellipse.
}
  \label{fig1}
\end{figure}

In Fig. \ref{fig1}, we display a photonic array of straight waveguides with honeycomb structure, consisting of two sublattices A and B.
The refractive index modulation depths ($\delta n_{\rm A}$ and $\delta n_{\rm B}$) in two sublattices can be made slightly different (detuned), as shown by different colors in Fig. \ref{fig1}.
This results in the breakup of the inversion symmetry of the array, disappearance of the Dirac cones in the spectrum, and opening of the gap between them.
It should be stressed that even though forbidden gap emerges in the band structure, the Berry curvature $\Omega$ of the first and second bulk bands satisfies the condition $\Omega(-{\bf k})=-\Omega({\bf k})$, which indicates that the Chern numbers of the two upper bulk bands remain zero \cite{xiao.rmp.82.1959.2010}. In this case six valleys appear in the spectrum, with three valleys around $\textbf{K}$ ($\textbf{K}'$) points being equivalent. Valley Chern number of a specific valley is determined to be either $+1/2$ or $-1/2$.
Moreover, if the valley Chern number for a certain valley is $+1/2$ for the lattice with $\delta n_{\rm A}>\delta n_{\rm B}$, then for the lattice with $\delta n_{\rm A}<\delta n_{\rm B}$ the Chern number for the same valley will be equal to $-1/2$ \cite{noh.prl.120.063902.2018}. Thus, if one designs a composite honeycomb lattice with a domain wall (highlighted by the red ellipse in Fig. \ref{fig1}) between two inversion-symmetry-broken honeycomb lattices with opposite detunings,
the valley Chern numbers at both sides of the interface become opposite.
In this case, bulk-edge correspondence principle, applied to valley Hall system, predicts the formation of the edge states localized on the domain wall and decaying in the direction perpendicular to it.
The appearance of such edge states is a manifestation of the well-known valley Hall effect \cite{mak.science.344.1489.2014,liu.aipx.6.1905546.2021,xue.apr.2.2100013.2021}
and corresponding edge states of topological origin are usually called valley Hall edge states \cite{wu.nc.8.1304.2017,noh.prl.120.063902.2018,shalaev.nn.14.31.2019,yang.np.14.446.2020}.

\section{Band structure and linear valley Hall edge state}

The propagation of the valley Hall edge state along the longitudinal $z$ axis of the waveguide array with focusing cubic nonlinearity can be described by the nonlinear Schr\"odinger equation,
\begin{equation}\label{eq1}
  i\frac{\partial \psi}{\partial z} = -\frac{1}{2} \left( \frac{\partial^2}{\partial x^2} + \frac{\partial^2}{\partial y^2} \right) \psi -{\mathcal R}(x,y) \psi - |\psi|^2 \psi,
\end{equation}
where $\psi$ is the dimensionless field amplitude, $x$ and $y$ are the normalized transverse coordinates, and $z$ is the normalized propagation distance, the function $\cal R$ stands for the refractive index distribution in the honeycomb array that is independent of the longitudinal coordinate $z$. The profiles of individual waveguides in the array can be described by Gaussian functions of width $\sigma$: ${\mathcal R}(x,y) = p_{\rm A,B} \sum_{m,n} e^{-[(x-x_{m,n})^2+(y-y_{m,n})^2]/\sigma^2}$, where $p_{\rm A,B}\sim \delta n_\textrm{A,B}$ stand for the depths of waveguides in two sublattices, and $(x_{m,n},y_{m,n})$ are the coordinates of the nodes in the honeycomb grid. We consider a configuration that is periodic along the $y$ axis and is limited along the $x$-axis with outer boundaries located far away from the domain wall, so that ${\mathcal R}(x,y)={\mathcal R}(x,y+L)$ with $L=\sqrt{3} a$ and $a$ being the array constant. As a representative parameter values we choose $a=1.4$ and $\sigma=0.5$. The average refractive index modulation depth is set to be $p_{\rm in}=10.3$, while the detuning $\delta=0.55$. For the honeycomb array on the left side of the domain wall in Fig. \ref{fig1} we set $p_{\rm A}=p_{\rm in}+\delta$ and $p_{\rm B}=p_{\rm in}-\delta$, while for the array on the right side of the domain wall we assume inverted detuning, so that $p_{\rm A}=p_{\rm in}-\delta$ and $p_{\rm B}=p_{\rm in}+\delta$. The domain wall emerging between these two arrays that we consider here is characterized by the reduced refractive index for all sites, see red ellipse in Fig. \ref{fig1}.
Normalized parameters described above correspond to the following real physical values in waveguide arrays inscribed in fused silica with femtosecond laser pulses
\cite{rechtsman.nature.496.196.2013,stuetzer.nature.560.461.2018,mukherjee.science.368.856.2020,maczewsky.science.370.701.2020,kirsch.np.2021,tan.ap.3.024002.2021} if laser radiation at the wavelength of $800~\rm nm$ is used and characteristic transverse scale is set to $10~\mu \rm m$, that corresponds to dimensionless coordinates $x,y=1$. In this case the array constant is $14~\mu \rm m$, waveguide width is $5~\mu \rm m$, and $p_{\rm in}=10.3$ corresponds to the refractive index modulation depth of $\sim1.1\times 10^{-3}$.

\begin{figure}[htpb]
\centering
\includegraphics[width=0.6\columnwidth]{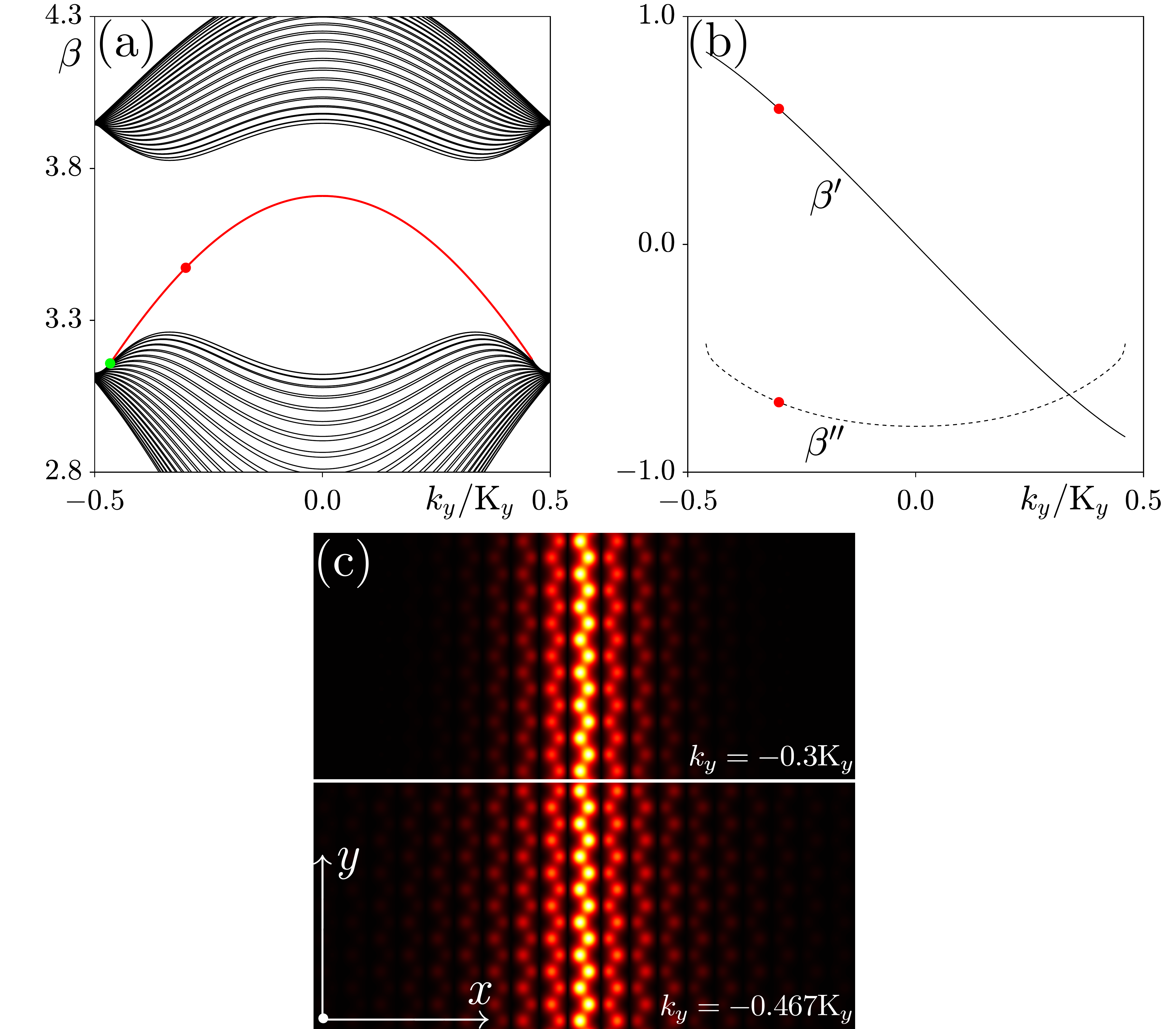}
\caption{(a) Band structure of the photonic graphene with a domain wall. The black curves are the bulk states, while the red curve is the valley Hall edge state. (b) First-order $\beta'$ and second-order $\beta''$ derivatives of the propagation constant of the valley Hall edge state. (c) Exemplary profiles of the valley Hall edge states with the Bloch momenta displayed in the right-bottom corner. These profiles correspond to the red and green dots in (a). The states are shown within the window $-20\le x \le 20$ and $-9.1\le y\le 9.1$.}
  \label{fig2}
\end{figure}

We obtained the bandgap structure of the composite array with a domain wall by substituting the solution $\psi(x,y,z) = u(x,y) e^{ik_y y + i\beta z}$ into linear counterpart of Eq. (\ref{eq1}).
Here $u(x,y)=u(x,y+L)$ is the periodic Bloch wave function, $k_y\in [-{\rm K}_y/2,{\rm K}_y/2)$ is the Bloch momentum in the first Brillouin zone with ${\rm K}_y=2\pi/L$,
and $\beta$ is the propagation constant of the linear mode that is a function of $k_y$. Using plane-wave expansion method we obtained the bandgap structure shown in Fig. \ref{fig2}(a),
in which the bulk states are indicated by the black lines and the in-gap valley Hall edge state is indicated by the red line.
To better understand properties of the edge state, we also display the first-order $\beta'=d\beta/dk_y$
(solid line) and second-order $\beta''=d^2\beta/dk_y^2$ (dashed line) derivatives of the propagation constant of the edge state in Fig. \ref{fig2}(b).
The first-order derivative $\beta'$ provides the group velocity $v=-\beta'$ with which edge state moves along the domain wall,
while the second-order derivative $\beta''$ quantifies the dispersion of the edge state and allows to estimate, in particular,
the rate of expansion along the domain wall of the localized envelope, if it is superimposed on the edge state.
As shown in Fig. \ref{fig2}(b), $\beta''$ is negative in the entire Brillouin zone which is necessary to obtain bright solitons. 
If the value is positive, one obtains dark solitons \cite{ren.nono.10.3559.2021}.
In Fig. \ref{fig2}(c), we display two examples of the linear valley Hall edge states corresponding to the red and greed dots in Fig. \ref{fig2}(a).
The localization of the state at $k_y=-0.3{\rm K}_y$ with propagation constant closer to the center of the gap (red dot) is much better than that of the state at $k_y=-0.467{\rm K}_y$ taken close to the gap edge (green dot). For both these $k_y$ values $\beta'>0$ that corresponds to the motion in the negative $y$ direction during propagation.
Notice that the same domain wall supports states propagating in the opposite direction, since the valley Hall system is time-reversal symmetric, and there must be a back-propagating state as the time-reversal conjugate of the forward-propagating state. In the valley Hall system, the counter-propagating states can hardly couple (such coupling is only possible under the action of strong localized defects that couple two valleys, while smooth large-scale perturbations do not couple them) making such systems beneficial in comparison with topologically trivial waveguide arrays. In the following we consider states with Bloch momentum $k_y=-0.3{\rm K}_y$, but point out that their properties remain similar for other $k_y$ values.

\section{Nonlinear valley Hall edge state and quasi-soliton}

\begin{figure}[htpb]
\centering
\includegraphics[width=0.6\columnwidth]{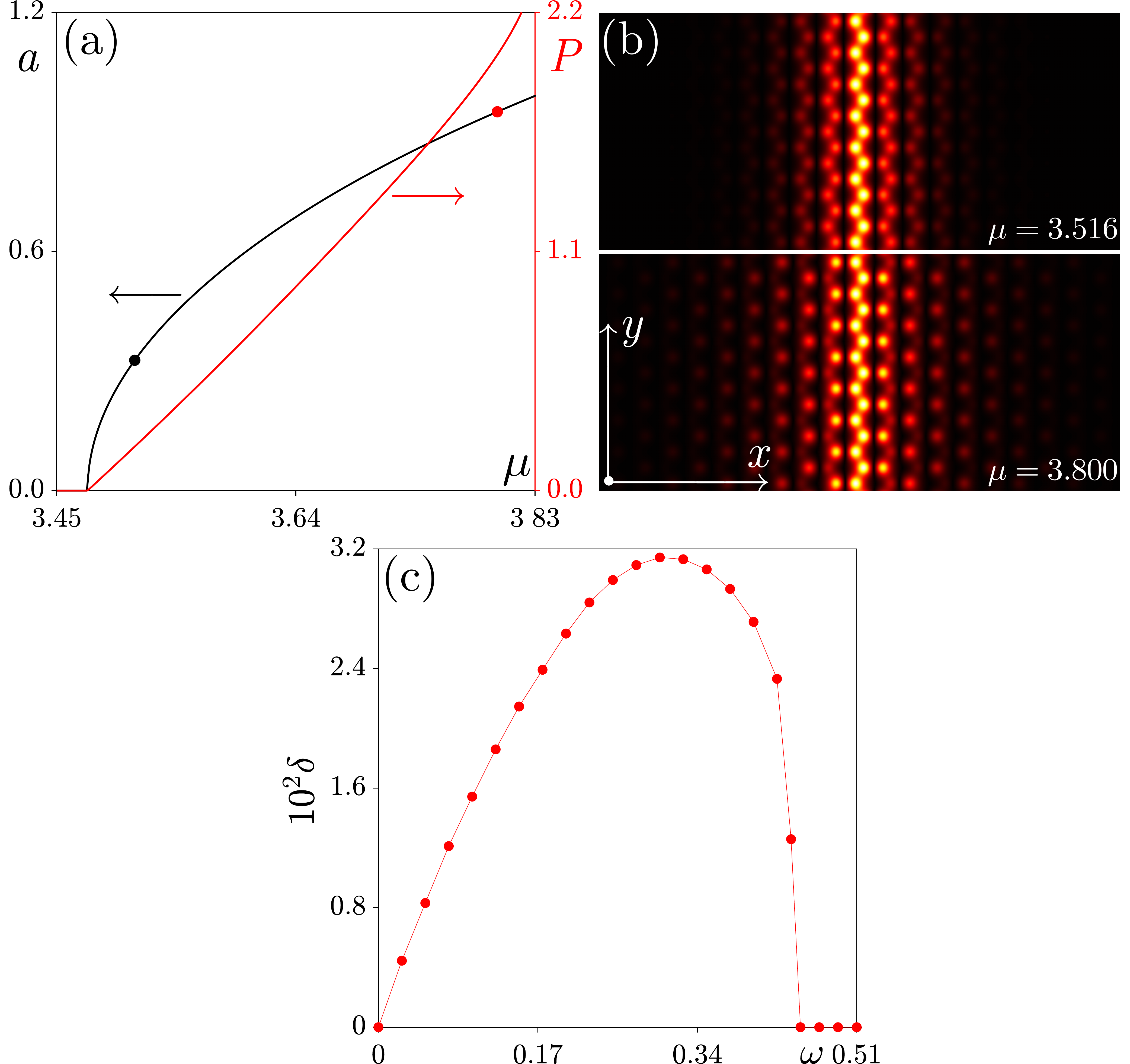}
\caption{(a) Peak amplitude (black curve; left vertical axis) and power (red curve; right vertical axis) of the nonlinear valley Hall edge state as a function of nonlinear propagation constant shift $\mu$. (b) Examples of profiles of the nonlinear valley Hall edge states for $\mu$ values displayed in the right-bottom corner and corresponding to the black and red dots in (a). The states are shown within the window $-20\le x \le 20$ and $-9.1\le y\le 9.1$.
(c) Growth rate $\delta$ of the small perturbation added to the nonlinear edge state with $\mu=3.516$ versus
frequency of the perturbation $\omega$.}
  \label{fig3}
\end{figure}

To obtain bright valley Hall edge solitons, we first calculate nonlinear extension of the valley Hall edge states. To do this, we insert the ansatz $\psi(x,y,z) = u(x,y) e^{ik_y y + i\mu z}$, where $\mu$ is the nonlinear propagation constant shift, into nonlinear Eq. (\ref{eq1}) that yields the equation
\begin{equation}\label{eq2}
  \mu u=\frac{1}{2}\left(\frac{\partial^{2}}{\partial x^{2}}+\frac{\partial^{2}}{\partial y^{2}}+2 i k_{y} \frac{\partial}{\partial y}-k_{y}^{2}\right) u+\mathcal{R}(x, y) u+|u|^{2} u,
\end{equation}
which can be solved by using Newton method for a given nonlinear propagation constant shift $\mu$ that lies in the interval $\beta_{\rm li} \le \mu \le \beta_{\rm ge}$. Here, $\beta_{\rm li}\approx3.473$ is the propagation constant of the linear valley Hall edge state at $k_y=-0.3{\rm K}_y$ [see the red dot in Fig. \ref{fig2}(a)], while $\beta_{\rm ge}\approx3.832$ is the propagation constant corresponding to the top edge of the gap for the same Bloch momentum $k_y$. The peak amplitude $a$ (solid curve) and power per one $y$-period of the structure $P=\int_{-\infty}^{+\infty}\int_0^L |\psi|^2 dx dy$ (red curve) for the nonlinear valley Hall edge state family are shown in Fig. \ref{fig3}(a). They both monotonously increase with increasing nonlinear propagation constant shift $\mu$ and they vanish exactly in the point, where nonlinear edge state family bifurcates from the linear family. Amplitude profiles $|u|$ of two representative nonlinear edge states with $\mu=3.516$ and $\mu=3.8$, corresponding respectively to the black and red dots in Fig. \ref{fig3}(a), are shown in Fig. \ref{fig3}(b). Since the state indicated by the red dot is much closer to the top edge of the gap, its localization is worse than that of the state corresponding to the black dot.
We choose the state corresponding to the black dot with $\mu=3.516$ and investigate its modulational instability by adding
a perturbation into its initial profile that is $\sim\nu\cos(\omega y)$, with $\nu=0.01$ and $\omega$ being the amplitude and the frequency, respectively. Such small perturbations experience clear exponential growth at the initial stage of instability development as long as modulation frequency $\omega$ is within the modulational instability band. The dependence of the perturbation growth rate $\delta$ on frequency $\omega$ can be easily obtained from direct simulations of propagation, as shown in Fig. \ref{fig3}(c). The dependence $\delta(\omega)$ unveils that the modulation instability bandwidth is finite.

\begin{figure}[htpb]
\centering
\includegraphics[width=0.6\columnwidth]{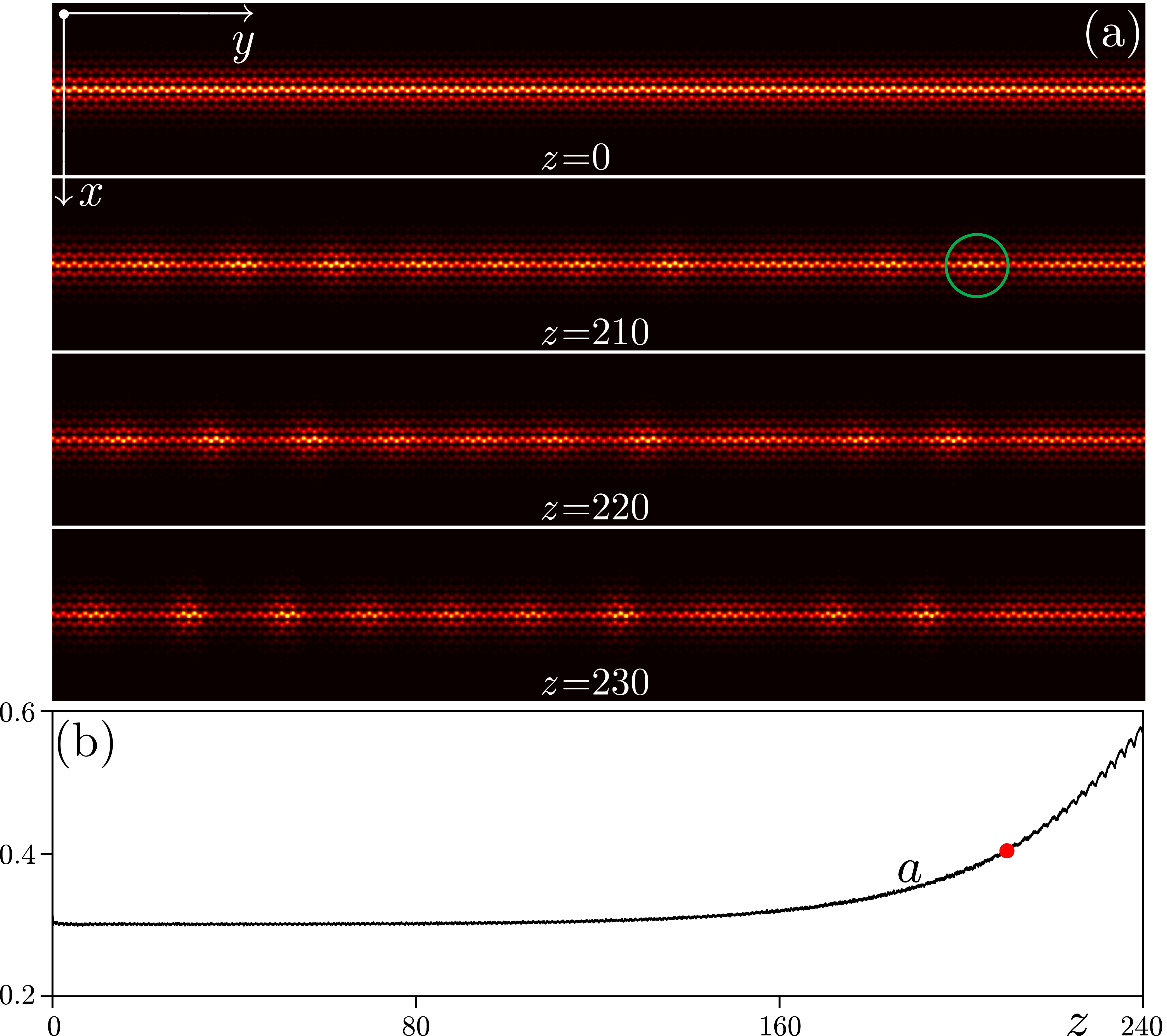}
\caption{(a) Amplitude profiles $|\psi|$ with $\mu=3.516$ of the perturbed propagating nonlinear valley Hall edge state at different propagation distances. (b) Peak amplitude $a$ of the state versus propagation distance. All states are shown within the window $-20\le x \le 20$ and $-121.2\le y\le 121.2$.}
\label{fig4}
\end{figure}

Next we consider dynamics of propagation of the nonlinear edge states. We are interested mostly in the edge states with not too small peak amplitudes (or their behavior will be close to that of linear edge states) that exhibit relatively fast decay in the course of propagation due to the development of modulational instability.
For example, one can choose the same nonlinear edge state, whose modulation instability is studied in Fig. \ref{fig3}(c).
To illustrate the development of modulational instability, one can introduce the periodic perturbation as adopted in Fig. \ref{fig3}(c).
Besides, one can also 
perturb the nonlinear valley Hall edge state by a random $5\%$-amplitude noise i.e., consider input in the form $\psi(x,y)[1+\delta(x,y)]$, where $\delta(x,y)$ is a random number uniformly distributed within the segment $[-0.05,+0.05]$ and propagate it in the structure containing $200$ $y$-periods, as shown in Fig. \ref{fig4}(a). The amplitude profiles shown at different propagation distances reveal the development of modulational instability,
which results in breakup of the wave into multiple bright spots --- precursors of bright solitons,
whose formation is possible in this system due to focusing nonlinearity and appropriate sign of the second-order dispersion $\beta''$.
Notice that instability development does not lead to dramatic radiation into the bulk, i.e. nearly all power remains in the vicinity of the domain wall.
The peak amplitude $a$ of the nonlinear state during propagation is depicted in Fig. \ref{fig4}(b).
The red dot on this dependence corresponds to $z=210$ and to sufficiently pronounced edge state modulations, as it follows from Fig. \ref{fig4}(a).

\begin{figure}[htpb]
\centering
\includegraphics[width=0.6\columnwidth]{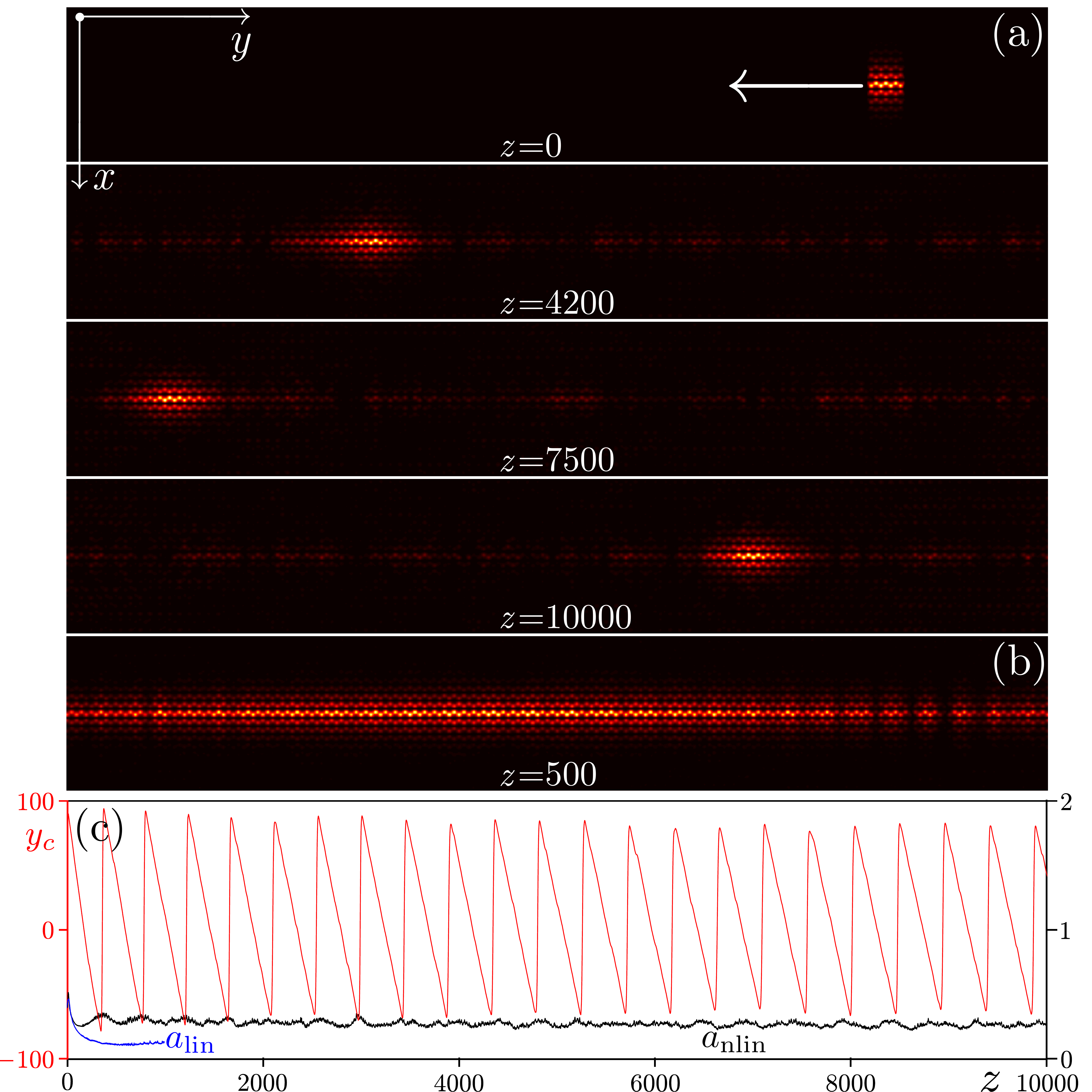}
\caption{(a) Long-range stable propagation dynamics of the valley Hall edge quasi-soliton. (b) Profile for the same input after linear propagation at $z=500$. (c) Peak amplitude (right axis) for nonlinear $a_{\rm nlin}$ (black curve) and linear $a_{\rm lin}$ (blue curve) propagation regimes and integral soliton center position $y_c$ in nonlinear regime (left axis) versus propagation distance $z$.
All amplitude distributions in (a) are shown within the window $-20\le x \le 20$ and $-121.2\le y\le 121.2$.}
\label{fig5}
\end{figure}

To confirm that isolated bright spots emerging as a result of modulation instability development indeed can give rise to stable valley Hall edge quasi-solitons 
(here ``quasi'' means that such states still exhibit small radiative losses during propagation, even though these losses are so weak that they do not lead to noticeable decrease of peak amplitude even at $z \sim 10^4$, see below), 
we selected one of such spots indicated by the green circle in Fig. \ref{fig4}(a) as an input state at $z=0$ in Fig. \ref{fig5}(a) and propagated it up to $z=10^4$. 
The evolution of the peak amplitude $a_{\rm nlin}$ (black curve) and integral center position $y_c = (\iint |\psi|^2 dx dy)^{-1} \iint y|\psi|^2 dx dy$ of the emerged quasi-soliton during propagation is presented in Fig. \ref{fig5}(c). 
One can see that after slight initial decrease, the peak amplitude $a_{\rm nlin}$ of so constructed input exhibits only small oscillations and does not decrease with distance, 
clearly indicating on the fact that nonlinear self-action has compensated diffraction broadening for this self-sustained state. 
The edge soliton moves along the $y$-axis in its negative direction with constant velocity and in our case traverses $y$-window (where we used periodic boundary conditions) multiple times, without any signature of diffractive broadening. 
In Fig. \ref{fig5}(a), we also display amplitude profiles of the quasi-soliton at different distances: they show that the profile of the quasi-soliton remains nearly unchanged. 
We would like to note that the input spot in Fig. \ref{fig5}(a) is not exactly the valley Hall edge soliton solution and 
this is the reason for slight initial decrease of the peak amplitude that corresponds to the stage at which wavepacket self-adjusts to soliton shape. 
If nonlinearity in Eq. (\ref{eq1}) is switched off, the same input quickly and dramatically spreads in linear medium, extending along the domain wall.
This is illustrated in Fig. \ref{fig5}(b), where we show the output distribution at $z=500$ after linear propagation, when it substantially extends along the domain wall. In Fig. \ref{fig5}(c), we also show the corresponding peak amplitude $a_{\rm lin}$ during linear propagation, but only within the region $z\le1000$.
Further propagation will lead to the interference of the state due to the limited size of the calculation window which affects the peak amplitude of the state.
Besides the method of generation of quasi-solitons adopted here that utilizes modulation instability \cite{kartashov.optica.3.1228.2016,zhang.pra.99.053836.2019,zhong.ap.3.056001.2021}, 
one can also derive an envelope equation for such solitons directly from Eq. (\ref{eq1}) using the methods developed for continuous topological systems in \cite{ivanov.acs.7.735.2020, ivanov.pra.103.053507.2021}.

\section{Topological protection of the valley Hall edge soliton}

\begin{figure}[htpb]
\centering
\includegraphics[width=\columnwidth]{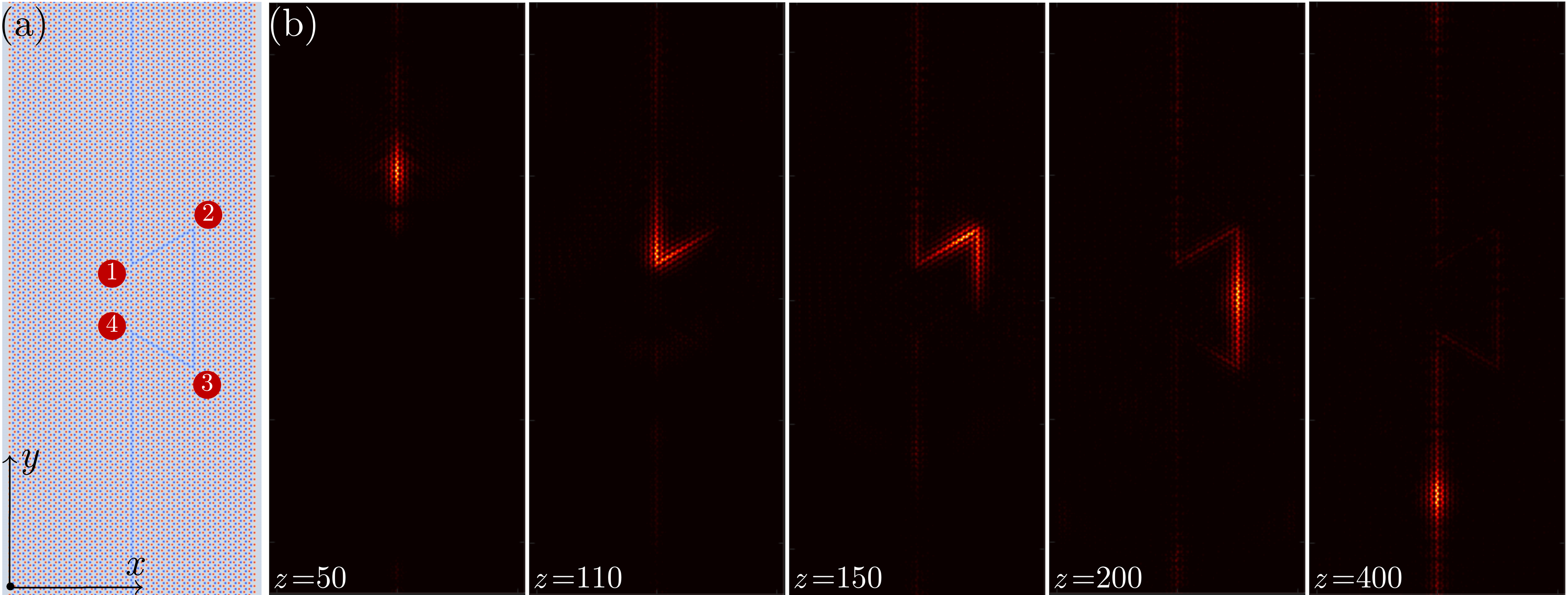}
\caption{(a) Inversion-symmetry-broken honeycomb lattice with a $\Omega$-shaped domain wall (the blue channel). We indicate four corners of this structure with numbers. (b) Amplitude profiles at different propagation distances illustrating passage of soliton through all sharp bends of the domain wall. All states are shown within the window $-53\le x \le 53$ and $-121.2\le y\le 121.2$.}
\label{fig6}
\end{figure}

One of the most representative properties of the topological edge states is their topological protection. 
While certain small-scale modulations of the domain wall may still cause backscattering in the valley Hall system, topological solitons in this system can circumvent sharp corners without backward reflection or radiation. 
Notice that in previously reported example \cite{zhong.ap.3.056001.2021} of the nonlinear valley Hall system with type-II Dirac cones, the specific geometry of the interface did not allow to illustrate this type of dynamics. 
In contrast, topological protection can be easily visualized in our system. We thus construct a domain wall with a $\Omega$-like shape, that possesses 4 sharp corners (the angle is $60^\circ$), as shown by the blue channel in Fig. \ref{fig6}(a). 
Since the lattice unit cell used in this work has $\rm C_{3v}$ symmetry, the formation of such zigzag-type turns with an angle of $60^\circ$ or $120^\circ$ is allowed.
We use the same input as in Fig. \ref{fig5}(a) and check its propagation dynamics along the $\Omega$-shaped domain wall, see Fig. \ref{fig6}(b). 
Presented results clearly show that soliton passes all sharp corners in the domain wall without experiencing reflection. 
Notice that one of the main advantages of our system is considerable width of the gap -- 
hence all obtained solitons have sufficiently large propagation velocities $v=-\beta'$ allowing them to pass through bends and corners over sufficiently small propagation distances 
[thus, $z=200$ in Fig. \ref{fig6}(b) is of the order of experimentally available sample length for laser-written waveguide arrays].
An animation corresponding to the propagation in Fig. \ref{fig6} is provided in the \textbf{Visualisation 1},
that visually shows the topological protection.

\section{Conclusion}

Summarizing, we have demonstrated valley Hall edge solitons in a composite honeycomb lattice with broken inversion symmetry.
We have shown that a domain wall created in the composite honeycomb lattice supports edge states originating from the valley Hall effect.
Their nonlinear counterparts bifurcating from the linear valley Hall edge states were obtained by using the Newton method.
We used modulational instability to demonstrate that such nonlinear edge states split into sets of solitons,
each of which can show extremely long stable propagation along the domain wall.
Finally, topological protection was illustrated by considering interactions of valley Hall edge solitons with $\Omega$-shaped domain walls.
Our work suggests experimentally feasible approach to generation of topological edge solitons that does not rely on longitudinal array modulations leading to enhanced losses.
Our results may be generalized to other platforms where nontrivial topology can be combined with nonlinear response of the system \cite{zhai.njp.18.080201.2016,rudner.nrp.2.229.2020,Kartashov.nrp.1.185.20219,malomed.rjp.64.106.2019,mihalache.rrp.73.403.2021}.

\begin{backmatter}
\bmsection{Funding}
National Natural Science Foundation of China (12074308, U1537210);
Russian Science Foundation (21-12-00096);
Fundamental Research Funds for the Central Universities (xzy012019038).

\bmsection{Disclosures}
The authors declare no conflicts of interest.

\bmsection{Data Availability Statement}
Data underlying the results presented in this paper are not publicly available at this time but may be obtained from the authors upon reasonable request.

\end{backmatter}

\bibliography{my_refs_library}

\begin{thebibliography}{10}
\newcommand{\enquote}[1]{``#1''}

\bibitem{hasan.rmp.82.3045.2010}
M.~Z. Hasan and C.~L. Kane, \enquote{Colloquium: Topological insulators,}
  {\protect\JournalTitle{Rev. Mod. Phys.}} \textbf{82}, 3045--3067 (2010).

\bibitem{qi.rmp.83.1057.2011}
X.-L. Qi and S.-C. Zhang, \enquote{Topological insulators and superconductors,}
  {\protect\JournalTitle{Rev. Mod. Phys.}} \textbf{83}, 1057--1110 (2011).

\bibitem{rechtsman.nature.496.196.2013}
M.~C. Rechtsman, J.~M. Zeuner, Y.~Plotnik, Y.~Lumer, D.~Podolsky, F.~Dreisow,
  S.~Nolte, M.~Segev, and A.~Szameit, \enquote{Photonic {F}loquet topological
  insulators,} {\protect\JournalTitle{Nature}} \textbf{496}, 196--200 (2013).

\bibitem{haldane.prl.100.013904.2008}
F.~D.~M. Haldane and S.~Raghu, \enquote{Possible realization of directional
  optical waveguides in photonic crystals with broken time-reversal symmetry,}
  {\protect\JournalTitle{Phys. Rev. Lett.}} \textbf{100}, 013904 (2008).

\bibitem{wang.nature.461.772.2009}
Z.~Wang, Y.~Chong, J.~D. Joannopoulos, and M.~Solja{\v{c}}i{\'c},
  \enquote{{Observation of unidirectional backscattering-immune topological
  electromagnetic states},} {\protect\JournalTitle{Nature}} \textbf{461},
  772--775 (2009).

\bibitem{lindner.np.7.490.2011}
N.~H. Lindner, G.~Refael, and V.~Galitski, \enquote{Floquet topological
  insulator in semiconductor quantum wells,} {\protect\JournalTitle{Nat.
  Phys.}} \textbf{7}, 490--495 (2011).

\bibitem{hafezi.np.7.907.2011}
M.~Hafezi, E.~A. Demler, M.~D. Lukin, and J.~M. Taylor, \enquote{Robust optical
  delay lines with topological protection,} {\protect\JournalTitle{Nat. Phys.}}
  \textbf{7}, 907--912 (2011).

\bibitem{stuetzer.nature.560.461.2018}
S.~St\"utzer, Y.~Plotnik, Y.~Lumer, P.~Titum, N.~H. Lindner, M.~Segev, M.~C.
  Rechtsman, and A.~Szameit, \enquote{Photonic topological {Anderson}
  insulators,} {\protect\JournalTitle{Nature}} \textbf{560}, 461--465 (2018).

\bibitem{yang.nature.565.622.2019}
Y.~Yang, Z.~Gao, H.~Xue, L.~Zhang, M.~He, Z.~Yang, R.~Singh, Y.~Chong,
  B.~Zhang, and H.~Chen, \enquote{Realization of a three-dimensional photonic
  topological insulator,} {\protect\JournalTitle{Nature}} \textbf{565},
  622--626 (2019).

\bibitem{mukherjee.science.368.856.2020}
S.~Mukherjee and M.~C. Rechtsman, \enquote{Observation of {F}loquet solitons in
  a topological bandgap,} {\protect\JournalTitle{Science}} \textbf{368},
  856--859 (2020).

\bibitem{maczewsky.science.370.701.2020}
L.~J. Maczewsky, M.~Heinrich, M.~Kremer, S.~K. Ivanov, M.~Ehrhardt,
  F.~Martinez, Y.~V. Kartashov, V.~V. Konotop, L.~Torner, D.~Bauer, and
  A.~Szameit, \enquote{Nonlinearity-induced photonic topological insulator,}
  {\protect\JournalTitle{Science}} \textbf{370}, 701--704 (2020).

\bibitem{yang.light.9.128.2020}
Z.~Yang, E.~Lustig, Y.~Lumer, and M.~Segev, \enquote{Photonic {F}loquet
  topological insulators in a fractal lattice,} {\protect\JournalTitle{Light
  Sci. Appl.}} \textbf{9}, 128 (2020).

\bibitem{yang.prl.114.114301.2015}
Z.~Yang, F.~Gao, X.~Shi, X.~Lin, Z.~Gao, Y.~Chong, and B.~Zhang,
  \enquote{Topological acoustics,} {\protect\JournalTitle{Phys. Rev. Lett.}}
  \textbf{114}, 114301 (2015).

\bibitem{peng.nc.7.13368.2016}
Y.-G. Peng, C.-Z. Qin, D.-G. Zhao, Y.-X. Shen, X.-Y. Xu, M.~Bao, H.~Jia, and
  X.-F. Zhu, \enquote{Experimental demonstration of anomalous {F}loquet
  topological insulator for sound,} {\protect\JournalTitle{Nat. Commun.}}
  \textbf{7}, 13368 (2016).

\bibitem{he.np.12.1124.2016}
C.~He, X.~Ni, H.~Ge, X.-C. Sun, Y.-B. Chen, M.-H. Lu, X.-P. Liu, and Y.-F.
  Chen, \enquote{Acoustic topological insulator and robust one-way sound
  transport,} {\protect\JournalTitle{Nat. Phys.}} \textbf{12}, 1124--1129
  (2016).

\bibitem{lu.np.13.369.2017}
J.~Lu, C.~Qiu, L.~Ye, X.~Fan, M.~Ke, F.~Zhang, and Z.~Liu, \enquote{Observation
  of topological valley transport of sound in sonic crystals,}
  {\protect\JournalTitle{Nat. Phys.}} \textbf{13}, 369--374 (2017).

\bibitem{zhang.cp.1.97.2018}
X.~Zhang, M.~Xiao, Y.~Cheng, M.-H. Lu, and J.~Christensen, \enquote{Topological
  sound,} {\protect\JournalTitle{Commun. Phys.}} \textbf{1}, 97 (2018).

\bibitem{ma.nrp.1.281.2019}
G.~Ma, M.~Xiao, and C.~T. Chan, \enquote{Topological phases in acoustic and
  mechanical systems,} {\protect\JournalTitle{Nat. Rev. Phys.}} \textbf{1},
  281--294 (2019).

\bibitem{susstrunk.science.349.47.2015}
R.~S{\"u}sstrunk and S.~D. Huber, \enquote{Observation of phononic helical edge
  states in a mechanical topological insulator,}
  {\protect\JournalTitle{Science}} \textbf{349}, 47--50 (2015).

\bibitem{huber.np.12.621.2016}
S.~D. Huber, \enquote{Topological mechanics,} {\protect\JournalTitle{Nat.
  Phys.}} \textbf{12}, 621--623 (2016).

\bibitem{goldman.pnas.110.6736.2013}
N.~Goldman, J.~Dalibard, A.~Dauphin, F.~Gerbier, M.~Lewenstein, P.~Zoller, and
  I.~B. Spielman, \enquote{Direct imaging of topological edge states in
  cold-atom systems,} {\protect\JournalTitle{Proc. Natl. Acad. Sci.}}
  \textbf{110}, 6736--6741 (2013).

\bibitem{jotzu.nature.515.237.2014}
G.~Jotzu, M.~Messer, R.~Desbuquois, M.~Lebrat, T.~Uehlinger, D.~Greif, and
  T.~Esslinger, \enquote{Experimental realisation of the topological {H}aldane
  model,} {\protect\JournalTitle{Nature}} \textbf{515}, 237--240 (2014).

\bibitem{nalitov.prl.114.116401.2015}
A.~V. Nalitov, D.~D. Solnyshkov, and G.~Malpuech, \enquote{Polariton
  $\mathbb{Z}$ topological insulator,} {\protect\JournalTitle{Phys. Rev.
  Lett.}} \textbf{114}, 116401 (2015).

\bibitem{jean.np.11.651.2017}
P.~St-Jean, V.~Goblot, E.~Galopin, A.~Lema\^{\i}tre, T.~Ozawa, L.~Le~Gratiet,
  I.~Sagnes, J.~Bloch, and A.~Amo, \enquote{Lasing in topological edge states
  of a one-dimensional lattice,} {\protect\JournalTitle{Nat. Photon.}}
  \textbf{11}, 651--656 (2017).

\bibitem{klembt.nature.562.552.2018}
S.~Klembt, T.~H. Harder, O.~A. Egorov, K.~Winkler, R.~Ge, M.~A. Bandres,
  M.~Emmerling, L.~Worschech, T.~C.~H. Liew, M.~Segev, C.~Schneider, and
  S.~H\"{o}fling, \enquote{Exciton-polariton topological insulator,}
  {\protect\JournalTitle{Nature}} \textbf{562}, 552--556 (2018).

\bibitem{albert.prl.114.173902.2015}
V.~V. Albert, L.~I. Glazman, and L.~Jiang, \enquote{Topological properties of
  linear circuit lattices,} {\protect\JournalTitle{Phys. Rev. Lett.}}
  \textbf{114}, 173902 (2015).

\bibitem{hadad.ne.1.178.2018}
Y.~Hadad, J.~C. Soric, A.~B. Khanikaev, and A.~Al\`{u}, \enquote{Self-induced
  topological protection in nonlinear circuit arrays,}
  {\protect\JournalTitle{Nat. Electron.}} \textbf{1}, 178--182 (2018).

\bibitem{imhof.np.14.925.2018}
S.~Imhof, C.~Berger, F.~Bayer, J.~Brehm, L.~W. Molenkamp, T.~Kiessling,
  F.~Schindler, C.~H. Lee, M.~Greiter, T.~Neupert, and R.~Thomale,
  \enquote{Topolectrical-circuit realization of topological corner modes,}
  {\protect\JournalTitle{Nature Physics}} \textbf{14}, 925--929 (2018).

\bibitem{olekhno.nc.11.1436.2020}
N.~A. Olekhno, E.~I. Kretov, A.~A. Stepanenko, P.~A. Ivanova, V.~V. Yaroshenko,
  E.~M. Puhtina, D.~S. Filonov, B.~Cappello, L.~Matekovits, and M.~A. Gorlach,
  \enquote{Topological edge states of interacting photon pairs emulated in a
  topolectrical circuit,} {\protect\JournalTitle{Nat. Commun.}} \textbf{11},
  1436 (2020).

\bibitem{helbig.np.16.747.2020}
T.~Helbig, T.~Hofmann, S.~Imhof, M.~Abdelghany, T.~Kiessling, L.~W. Molenkamp,
  C.~H. Lee, A.~Szameit, M.~Greiter, and R.~Thomale, \enquote{Generalized
  bulk-boundary correspondence in non-{H}ermitian topolectrical circuits,}
  {\protect\JournalTitle{Nat. Phys.}} \textbf{16}, 747--750 (2020).

\bibitem{li.nsr.8.nwaa1192.2021}
R.~Li, B.~Lv, H.~Tao, J.~Shi, Y.~Chong, B.~Zhang, and H.~Chen, \enquote{Ideal
  type-{II} {Weyl} points in topological circuits,}
  {\protect\JournalTitle{National Sci. Rev.}} \textbf{8}, nwaa192 (2020).

\bibitem{lu.np.8.821.2014}
L.~Lu, J.~D. Joannopoulos, and M.~Solja{\v{c}}i{\'c}, \enquote{Topological
  photonics,} {\protect\JournalTitle{Nat. Photon.}} \textbf{8}, 821--829
  (2014).

\bibitem{ozawa.rmp.91.015006.2019}
T.~Ozawa, H.~M. Price, A.~Amo, N.~Goldman, M.~Hafezi, L.~Lu, M.~C. Rechtsman,
  D.~Schuster, J.~Simon, O.~Zilberberg, and I.~Carusotto, \enquote{Topological
  photonics,} {\protect\JournalTitle{Rev. Mod. Phys.}} \textbf{91}, 015006
  (2019).

\bibitem{kim.lsa.9.130.2020}
M.~Kim, Z.~Jacob, and J.~Rho, \enquote{Recent advances in 2{D}, 3{D} and
  higher-order topological photonics,} {\protect\JournalTitle{Light Sci.
  Appl.}} \textbf{9}, 130 (2020).

\bibitem{smirnova.apr.7.021306.2020}
D.~Smirnova, D.~Leykam, Y.~Chong, and Y.~Kivshar, \enquote{Nonlinear
  topological photonics,} {\protect\JournalTitle{Appl. Phys. Rev.}} \textbf{7},
  021306 (2020).

\bibitem{ota.nano.9.547.2020}
Y.~Ota, K.~Takata, T.~Ozawa, A.~Amo, Z.~Jia, B.~Kante, M.~Notomi, Y.~Arakawa,
  and S.~Iwamoto, \enquote{Active topological photonics,}
  {\protect\JournalTitle{Nanophoton.}} \textbf{9}, 547--567 (2020).

\bibitem{leykam.nano.9.4473.2020}
D.~Leykam and L.~Yuan, \enquote{Topological phases in ring resonators: recent
  progress and future prospects,} {\protect\JournalTitle{Nanophoton.}}
  \textbf{9}, 4473--4487 (2020).

\bibitem{segev.nano.10.425.2021}
M.~Segev and M.~A. Bandres, \enquote{Topological photonics: Where do we go from
  here?} {\protect\JournalTitle{Nanophoton.}} \textbf{10}, 425--434 (2021).

\bibitem{parto.nano.10.403.2021}
M.~Parto, Y.~G.~N. Liu, B.~Bahari, M.~Khajavikhan, and D.~N. Christodoulides,
  \enquote{Non-{H}ermitian and topological photonics: optics at an exceptional
  point,} {\protect\JournalTitle{Nanophoton.}} \textbf{10}, 403--423 (2021).

\bibitem{wang.fo.13.50.2020}
H.~Wang, S.~K. Gupta, B.~Xie, and M.~Lu, \enquote{Topological photonic
  crystals: a review,} {\protect\JournalTitle{Front. Optoelectron.}}
  \textbf{13}, 50--72 (2020).

\bibitem{liu.col.19.052602.2021}
H.~Liu, B.~Xie, H.~Cheng, J.~Tian, and S.~Chen, \enquote{Topological photonic
  states in artificial microstructures [{I}nvited],}
  {\protect\JournalTitle{Chin. Opt. Lett.}} \textbf{19}, 052602 (2021).

\bibitem{kartashov.prl.119.253904.2017}
Y.~V. Kartashov and D.~V. Skryabin, \enquote{Bistable topological insulator
  with exciton-polaritons,} {\protect\JournalTitle{Phys. Rev. Lett.}}
  \textbf{119}, 253904 (2017).

\bibitem{zhang.lpr.13.1900198.2019}
W.~Zhang, X.~Chen, Y.~V. Kartashov, D.~V. Skryabin, and F.~Ye,
  \enquote{Finite-dimensional bistable topological insulators: From small to
  large,} {\protect\JournalTitle{Laser Photon. Rev.}} \textbf{13}, 1900198
  (2019).

\bibitem{zhang.pra.99.053836.2019}
Y.~Q. Zhang, Y.~V. Kartashov, and A.~Ferrando, \enquote{Interface states in
  polariton topological insulators,} {\protect\JournalTitle{Phys. Rev. A}}
  \textbf{99}, 053836 (2019).

\bibitem{lumer.pra.94.021801.2016}
Y.~Lumer, M.~C. Rechtsman, Y.~Plotnik, and M.~Segev, \enquote{Instability of
  bosonic topological edge states in the presence of interactions,}
  {\protect\JournalTitle{Phys. Rev. A}} \textbf{94}, 021801 (2016).

\bibitem{kartashov.optica.3.1228.2016}
Y.~V. Kartashov and D.~V. Skryabin, \enquote{Modulational instability and
  solitary waves in polariton topological insulators,}
  {\protect\JournalTitle{Optica}} \textbf{3}, 1228--1236 (2016).

\bibitem{harari.science.359.eaar4003.2018}
G.~Harari, M.~A. Bandres, Y.~Lumer, M.~C. Rechtsman, Y.~D. Chong,
  M.~Khajavikhan, D.~N. Christodoulides, and M.~Segev, \enquote{Topological
  insulator laser: Theory,} {\protect\JournalTitle{Science}} \textbf{359},
  eaar4003 (2018).

\bibitem{bandres.science.359.eaar4005.2018}
M.~A. Bandres, S.~Wittek, G.~Harari, M.~Parto, J.~Ren, M.~Segev, D.~N.
  Christodoulides, and M.~Khajavikhan, \enquote{Topological insulator laser:
  Experiments,} {\protect\JournalTitle{Science}} \textbf{359}, eaar4005 (2018).

\bibitem{dikopoltsev.science.373.1514.2021}
A.~Dikopoltsev, T.~H. Harder, E.~Lustig, O.~A. Egorov, J.~Beierlein, A.~Wolf,
  Y.~Lumer, M.~Emmerling, C.~Schneider, S.~H\"ofling, M.~Segev, and S.~Klembt,
  \enquote{Topological insulator vertical-cavity laser array,}
  {\protect\JournalTitle{Science}} \textbf{373}, 1514--1517 (2021).

\bibitem{bahari.science.358.636.2017}
B.~Bahari, A.~Ndao, F.~Vallini, A.~El~Amili, Y.~Fainman, and B.~Kant{\'e},
  \enquote{Nonreciprocal lasing in topological cavities of arbitrary
  geometries,} {\protect\JournalTitle{Science}} \textbf{358}, 636--640 (2017).

\bibitem{kartashov.prl.122.083902.2019}
Y.~V. Kartashov and D.~V. Skryabin, \enquote{Two-dimensional topological
  polariton laser,} {\protect\JournalTitle{Phys. Rev. Lett.}} \textbf{122},
  083902 (2019).

\bibitem{zeng.nature.578.246.2020}
Y.~Zeng, U.~Chattopadhyay, B.~Zhu, B.~Qiang, J.~Li, Y.~Jin, L.~Li, A.~G.
  Davies, E.~H. Linfield, B.~Zhang, Y.~Chong, and Q.~J. Wang,
  \enquote{Electrically pumped topological laser with valley edge modes,}
  {\protect\JournalTitle{Nature}} \textbf{578}, 246--250 (2020).

\bibitem{zhong.lpr.14.2000001.2020}
H.~Zhong, Y.~D. Li, D.~H. Song, Y.~V. Kartashov, Y.~Q. Zhang, Y.~P. Zhang, and
  Z.~Chen, \enquote{Topological valley {H}all edge state lasing,}
  {\protect\JournalTitle{Laser Photon. Rev.}} \textbf{14}, 2000001 (2020).

\bibitem{gong.acs.7.2089.2020}
Y.~Gong, S.~Wong, A.~J. Bennett, D.~L. Huffaker, and S.~S. Oh,
  \enquote{Topological insulator laser using valley-hall photonic crystals,}
  {\protect\JournalTitle{ACS Photon.}} \textbf{7}, 2089--2097 (2020).

\bibitem{Kartashov.nrp.1.185.20219}
Y.~V. Kartashov, G.~E. Astrakharchik, B.~A. Malomed, and L.~Torner,
  \enquote{Frontiers in multidimensional self-trapping of nonlinear fields and
  matter,} {\protect\JournalTitle{Nat. Rev. Phys.}} \textbf{1}, 185--197
  (2019).

\bibitem{malomed.rjp.64.106.2019}
B.~A. Malomed and D.~Mihalache, \enquote{Nonlinear waves in optical and
  matter-wave media: A topical survey of recent theoretical and experimental
  results,} {\protect\JournalTitle{Rom. J. Phys.}} \textbf{64}, 106 (2019).

\bibitem{mihalache.rrp.73.403.2021}
D.~Mihalache, \enquote{Localized structures in optical and matter-wave media: A
  selection of recent studies,} {\protect\JournalTitle{Rom. Rep. Phys.}}
  \textbf{73}, 403 (2021).

\bibitem{lumer.prl.111.243905.2013}
Y.~Lumer, Y.~Plotnik, M.~C. Rechtsman, and M.~Segev, \enquote{Self-localized
  states in photonic topological insulators,} {\protect\JournalTitle{Phys. Rev.
  Lett.}} \textbf{111}, 243905 (2013).

\bibitem{bleu.nc.9.3991.2018}
O.~Bleu, G.~Malpuech, and D.~D. Solnyshkov, \enquote{Robust quantum valley
  {H}all effect for vortices in an interacting bosonic quantum fluid,}
  {\protect\JournalTitle{Nat. Commun.}} \textbf{9}, 3991 (2018).

\bibitem{leykam.prl.117.143901.2016}
D.~Leykam and Y.~D. Chong, \enquote{Edge solitons in nonlinear-photonic
  topological insulators,} {\protect\JournalTitle{Phys. Rev. Lett.}}
  \textbf{117}, 143901 (2016).

\bibitem{ablowitz.pra.96.043868.2017}
M.~J. Ablowitz and J.~T. Cole, \enquote{Tight-binding methods for general
  longitudinally driven photonic lattices: Edge states and solitons,}
  {\protect\JournalTitle{Phys. Rev. A}} \textbf{96}, 043868 (2017).

\bibitem{gulevich.sr.7.1780.2017}
D.~R. Gulevich, D.~Yudin, D.~V. Skryabin, I.~V. Iorsh, and I.~A. Shelykh,
  \enquote{Exploring nonlinear topological states of matter with
  exciton-polaritons: Edge solitons in kagome lattice,}
  {\protect\JournalTitle{Sci. Rep.}} \textbf{7}, 1780 (2017).

\bibitem{li.prb.97.081103.2018}
C.~Li, F.~Ye, X.~Chen, Y.~V. Kartashov, A.~Ferrando, L.~Torner, and D.~V.
  Skryabin, \enquote{Lieb polariton topological insulators,}
  {\protect\JournalTitle{Phys. Rev. B}} \textbf{97}, 081103 (2018).

\bibitem{smirnova.lpr.13.1900223.2019}
D.~A. Smirnova, L.~A. Smirnov, D.~Leykam, and Y.~S. Kivshar,
  \enquote{Topological edge states and gap solitons in the nonlinear {D}irac
  model,} {\protect\JournalTitle{Laser Photon. Rev.}} \textbf{13}, 1900223
  (2019).

\bibitem{zhang.prl.123.254103.2019}
W.~Zhang, X.~Chen, Y.~V. Kartashov, V.~V. Konotop, and F.~Ye, \enquote{Coupling
  of edge states and topological {B}ragg solitons,}
  {\protect\JournalTitle{Phys. Rev. Lett.}} \textbf{123}, 254103 (2019).

\bibitem{ivanov.acs.7.735.2020}
S.~K. Ivanov, Y.~V. Kartashov, A.~Szameit, L.~Torner, and V.~V. Konotop,
  \enquote{Vector topological edge solitons in {F}loquet insulators,}
  {\protect\JournalTitle{ACS Photon.}} \textbf{7}, 735--745 (2020).

\bibitem{ivanov.ol.45.1459.2020}
S.~K. Ivanov, Y.~V. Kartashov, L.~J. Maczewsky, A.~Szameit, and V.~V. Konotop,
  \enquote{Edge solitons in {L}ieb topological {F}loquet insulator,}
  {\protect\JournalTitle{Opt. Lett.}} \textbf{45}, 1459--1462 (2020).

\bibitem{ivanov.ol.45.2271.2020}
S.~K. Ivanov, Y.~V. Kartashov, L.~J. Maczewsky, A.~Szameit, and V.~V. Konotop,
  \enquote{Bragg solitons in topological {F}loquet insulators,}
  {\protect\JournalTitle{Opt. Lett.}} \textbf{45}, 2271--2274 (2020).

\bibitem{ivanov.pra.103.053507.2021}
S.~K. Ivanov, Y.~V. Kartashov, M.~Heinrich, A.~Szameit, L.~Torner, and V.~V.
  Konotop, \enquote{Topological dipole {F}loquet solitons,}
  {\protect\JournalTitle{Phys. Rev. A}} \textbf{103}, 053507 (2021).

\bibitem{zhong.ap.3.056001.2021}
H.~Zhong, S.~Xia, Y.~Zhang, Y.~Li, D.~Song, C.~Liu, and Z.~Chen,
  \enquote{{Nonlinear topological valley {Hall} edge states arising from
  type-{II} {Dirac} cones},} {\protect\JournalTitle{Adv. Photon.}} \textbf{3},
  056001 (2021).

\bibitem{smirnova.prr.3.043027.2021}
D.~A. Smirnova, L.~A. Smirnov, E.~O. Smolina, D.~G. Angelakis, and D.~Leykam,
  \enquote{Gradient catastrophe of nonlinear photonic valley-{Hall} edge
  pulses,} {\protect\JournalTitle{Phys. Rev. Research}} \textbf{3}, 043027
  (2021).

\bibitem{zangeneh.prl.123.053902.2019}
F.~Zangeneh-Nejad and R.~Fleury, \enquote{Nonlinear second-order topological
  insulators,} {\protect\JournalTitle{Phys. Rev. Lett.}} \textbf{123}, 053902
  (2019).

\bibitem{xiao.rmp.82.1959.2010}
D.~Xiao, M.-C. Chang, and Q.~Niu, \enquote{Berry phase effects on electronic
  properties,} {\protect\JournalTitle{Rev. Mod. Phys.}} \textbf{82}, 1959--2007
  (2010).

\bibitem{noh.prl.120.063902.2018}
J.~Noh, S.~Huang, K.~P. Chen, and M.~C. Rechtsman, \enquote{Observation of
  photonic topological valley {H}all edge states,} {\protect\JournalTitle{Phys.
  Rev. Lett.}} \textbf{120}, 063902 (2018).

\bibitem{mak.science.344.1489.2014}
K.~F. Mak, K.~L. McGill, J.~Park, and P.~L. McEuen, \enquote{The valley {H}all
  effect in {${\rm MoS}_2$} transistors,} {\protect\JournalTitle{Science}}
  \textbf{344}, 1489--1492 (2014).

\bibitem{liu.aipx.6.1905546.2021}
J.-W. Liu, F.-L. Shi, X.-T. He, G.-J. Tang, W.-J. Chen, X.-D. Chen, and J.-W.
  Dong, \enquote{Valley photonic crystals,} {\protect\JournalTitle{Adv. Phys.
  X}} \textbf{6}, 1905546 (2021).

\bibitem{xue.apr.2.2100013.2021}
H.~Xue, Y.~Yang, and B.~Zhang, \enquote{Topological valley photonics: Physics
  and device applications,} {\protect\JournalTitle{Adv. Photon. Research}}
  \textbf{2}, 2100013 (2021).

\bibitem{wu.nc.8.1304.2017}
X.~Wu, Y.~Meng, J.~Tian, Y.~Huang, H.~Xiang, D.~Han, and W.~Wen,
  \enquote{Direct observation of valley-polarized topological edge states in
  designer surface plasmon crystals,} {\protect\JournalTitle{Nat. Commun.}}
  \textbf{8}, 1304 (2017).

\bibitem{shalaev.nn.14.31.2019}
M.~I. Shalaev, W.~Walasik, A.~Tsukernik, Y.~Xu, and N.~M. Litchinitser,
  \enquote{Robust topologically protected transport in photonic crystals at
  telecommunication wavelengths,} {\protect\JournalTitle{Nat. Nanotech.}}
  \textbf{14}, 31--34 (2019).

\bibitem{yang.np.14.446.2020}
Y.~Yang, Y.~Yamagami, X.~Yu, P.~Pitchappa, J.~Webber, B.~Zhang, M.~Fujita,
  T.~Nagatsuma, and R.~Singh, \enquote{Terahertz topological photonics for
  on-chip communication,} {\protect\JournalTitle{Nat. Photon.}} \textbf{14},
  446--451 (2020).

\bibitem{kirsch.np.2021}
M.~S. Kirsch, Y.~Zhang, M.~Kremer, L.~J. Maczewsky, S.~K. Ivanov, Y.~V.
  Kartashov, L.~Torner, D.~Bauer, A.~Szameit, and M.~Heinrich,
  \enquote{Nonlinear second-order photonic topological insulators,}
  {\protect\JournalTitle{Nat. Phys.}} \textbf{17} (2021).

\bibitem{tan.ap.3.024002.2021}
D.~Tan, Z.~Wang, B.~Xu, and J.~Qiu, \enquote{Photonic circuits written by
  femtosecond laser in glass: improved fabrication and recent progress in
  photonic devices,} {\protect\JournalTitle{Adv. Photon.}} \textbf{3}, 024002
  (2021).

\bibitem{ren.nono.10.3559.2021}
B.~Ren, H.~Wang, V.~O. Kompanets, Y.~V. Kartashov, Y.~Li, and Y.~Zhang,
  \enquote{Dark topological valley {Hall} edge solitons,}
  {\protect\JournalTitle{Nanophoton.}} \textbf{10}, 3559--3566 (2021).

\bibitem{zhai.njp.18.080201.2016}
H.~Zhai, M.~Rechtsman, Y.-M. Lu, and K.~Yang, \enquote{Focus on topological
  physics: from condensed matter to cold atoms and optics,}
  {\protect\JournalTitle{New J. Phys.}} \textbf{18}, 080201 (2016).

\bibitem{rudner.nrp.2.229.2020}
M.~S. Rudner and N.~H. Lindner, \enquote{Band structure engineering and
  non-equilibrium dynamics in {F}loquet topological insulators,}
  {\protect\JournalTitle{Nat. Rev. Phys.}} \textbf{2}, 229--244 (2020).

\end{thebibliography}

\end{document}